\documentclass[graybox]{svmult}

\usepackage{subfigure}

\newcommand{\nc}{\newcommand}
\nc{\numberthis}{\addtocounter{equation}{1}\tag{\theequation}}

\nc{\dsp}[1]{^\mathrm{#1}}
\nc{\lb}{\left (}
\nc{\rb}{\right )}
\nc{\lset}{\left \{}
\nc{\rset}{\right \}}
\nc{\eqtext}[1]{\quad \text{#1} \quad}
\nc{\lsq}{\left [}
\nc{\rsq}{\right ]}

\newcommand{\dd}[1]{\; \mathrm{d} #1}

\newcommand{\sechn}[2]{\mathrm{sech}^{#1} \lb #2 \rb}
\newcommand{\tanhn}[2]{\mathrm{tanh}^{#1} \lb #2 \rb}

\usepackage{type1cm}        
\usepackage{makeidx}         
\usepackage{graphicx}        
\usepackage{multicol}        
\usepackage[bottom]{footmisc}

\usepackage{newtxtext}       %
\usepackage{newtxmath}       

\makeindex             

\begin{document}

\title*{Nonlinear Longitudinal Bulk Strain Waves in Layered Elastic Wavegudes}
\author{Karima R. Khusnutdinova and Matthew R. Tranter\\[5ex]
}
\institute{Karima R. Khusnutdinova (corresponding author) \at Department of Mathematical Sciences, Loughborough University, Loughborough LE11 3TU, UK \\ \email{K.Khusnutdinova@lboro.ac.uk} 
\and Matthew R. Tranter \at Department of Mathematical Sciences, Loughborough University, Loughborough LE11 3TU, UK\\  \email{M.R.Tranter@lboro.ac.uk}}
%
%
\maketitle

\abstract{We consider long longitudinal bulk strain waves in layered waveguides using Boussinesq-type equations. The equations are developed using lattice models, and this is viewed as an extension of the Fermi-Pasta-Ulam problem. We describe semi-analytical approaches to the solution of scattering problems in delaminated waveguides, and to the construction of the solution of an initial-value problem in the class of periodic functions, motivated by the scattering problems.}

\section{Introduction}
\label{sec:1}

Layered structures are frequently used in modern engineering constructions. The dynamical behaviour of layered structures depends not only on the properties of the bulk material, but also on the type of the bonding between the layers. For example, if layers have similar properties and the bonding between the layers is sufficiently soft, then the bulk strain soliton is replaced with a solitary wave radiating a co-propagating oscillatory tail \cite{KSZ}. Experimental observations of both pure and radiating bulk strain solitons in layered bars have been discussed in  \cite{Dreiden12}. 

The aim of this paper is to describe some efficient semi-analytical approaches to the modelling of the scattering of pure and radiating solitary waves in delaminated areas of bonded layered structures developed in \cite{KS, KT1, KT2}, as well as some initial-value problems motivated by these studies \cite{KM, KMP, KT3}. In Section 2, the model equations are discussed within the framework of lattice models which are presented as an extension of the famous Fermi-Pasta-Ulam (FPU) problem (also known as Fermi-Pasta-Ulam-Tsingou (FPUT) problem, see \cite{Dauxois}). Section 3 is devoted to the scattering of pure and radiating solitary waves in bi-layers with delamination. Section 4 is devoted to the discussion of the apparent zero-mass contradiction between the solutions of the original problem formulation and its weakly-nonlinear counterpart, which appears in periodic problems. We construct a solution of the initial-value problem which bypasses this difficulty. We conclude in Section 5.

\section{From Fermi-Pasta-Ulam chain to the model of a layered waveguide}
\label{sec:2}

\begin{figure}[!htbp]
	\centering
	\includegraphics[width=0.6\textwidth]{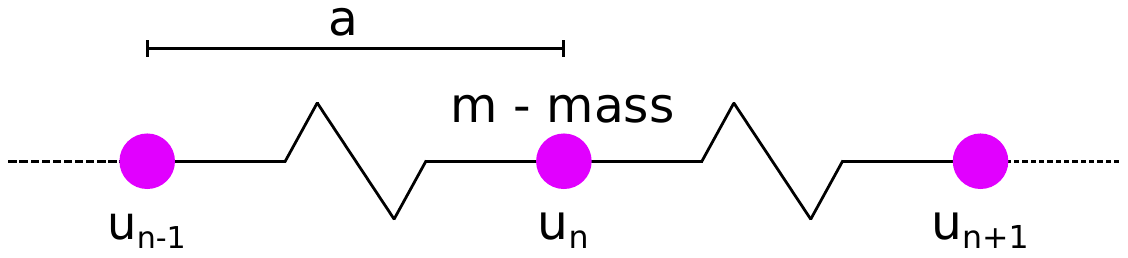}
	\caption{\small A schematic of an FPU chain with particles of mass $m$ and separation $a$, connected by weakly-nonlinear springs.}
	\label{fig:FPU}
\end{figure}

The FPU  problem is considered to be the starting point of modern nonlinear wave theory. In 1955, in Los Alamos, Fermi, Pasta and Ulam numerically studied the dynamics of an anharmonic chain of particles \cite{FPU} (see \cite{Dauxois} for the discussion of the contribution by Mary Tsingou).  The model consisted of identical equidistant (by a distance $a$) particles connected to their nearest neighbours by weakly-nonlinear springs (see Figure \ref{fig:FPU}).

Let $u_n$ denote the displacement of the $n$-th particle from the equilibrium. Then, using Newton's second law one obtains
\begin{equation*}
	m \ddot{u}_n = f(u_{n+1}-u_n) - f(u_n-u_{n-1}), \quad n=\overline{1, N},
\end{equation*}
where $f(\Delta u)$ was given by  $f(\Delta u) = k \Delta u + \alpha (\Delta u)^2$.
The ends of the chain were fixed: $u_0 = u_{N+1} = 0$. Thus, the dynamics of the chain was described by the system
\begin{align}
	&m \ddot{u}_n = k \lb u_{n+1} - 2 u_n + u_{n-1} \rb + \alpha \lsq \lb u_{n+1} - u_n \rb^2 - \lb u_n - u_{n-1} \rb^2 \rsq,\quad n = \overline{1,N}, \nonumber  \\
	&u_0 = u_{N+1} = 0 \label{1}
\end{align}
(typically, $N$ was equal to 64).

The general solution of the linearised system ($\alpha=0$) is given by a superposition of normal modes:
\begin{align*}
	u_n^k (t) &= A_k \sin \lb \frac{\pi k n }{N+1}  \rb \cos \ \lb \omega_k t + \delta_k \rb,\quad k = 1, \dots, N,  \\
	\omega_k &= 2 \sqrt{\frac km} \sin \left [ \frac{\pi k }{2 (N+1)} \right ],
\end{align*}
where $A_k$ and $\delta_k$ are arbitrary constants.
There is no energy transfer between the modes in the linear approximation.
In the nonlinear chain ($\alpha \ne 0$), modes become coupled. It was expected that if all the initial energy was put into a single mode (or a few of the first modes), the nonlinear coupling would yield equal distribution of the energy among the normal modes. However, the numerical results were surprising: for example, if the energy was initially in the mode of lowest frequency, it returned almost entirely to that mode after interaction with a few other low frequency modes  (FPU recurrence).

In 1965 this strange observation has motivated Zabusky and Kruskal to consider the FPU problem in the so-called continuum approximation \cite{ZK}. One assumes that
\begin{equation*}
	u_n (t) = u(x_n, t) = u(n a, t), \quad u_{n\pm1} = u(x_n \pm a, t),
\end{equation*}
and the displacement field $u$ varies slowly justifying the Taylor expansion
\begin{align}
	u_{n\pm1} (t) &\approx u(x_n, t) \pm a u' (x_n, t) + \frac{1}{2} a^2 u'' (x_n, t) \nonumber \\
	&\pm \frac{1}{6} a^3 u''' (x_n, t) + \frac{1}{24} a^4 u^{''''} (x_n, t) + \dots. \label{2}
\end{align}
Substituting (\ref{2}) into (\ref{1}) and dropping the label $n$ yields the equation
\begin{equation}
	u_{tt}-c^2 u_{xx} = \varepsilon c^2 (u_x u_{xx} + \delta^2 u_{xxxx}),
	\label{3}
\end{equation}
where $c^2 = \frac{ka^2}{m}, \varepsilon = \frac{2 \alpha a}{k}, \delta^2 = \frac{a^2}{12 \varepsilon}$. Here, the leading-order nonlinear and dispersive contributions are balanced at the same order of $\varepsilon$. This is the  Boussinesq equation. It describes waves, which can propagate both to the right, and to the left (the two-way long-wave equation).

A further reduction to the  Korteweg - de Vries (KdV) equation was obtained by using an asymptotic multiple-scales expansion of the solution of (\ref{3}). We assume a solution of the form
\begin{equation*}
	u = f(\xi, T) + \varepsilon u^{(1)} (x, t) + ..., \quad \mbox{where} \quad \xi = x-ct, \quad T = \varepsilon t,
\end{equation*}
then (\ref{3}) gives us
\begin{equation*}
	u^{(1)}_{tt} - c^2 u^{(1)}_{xx} = 2 c f_{\xi T} + c^2 f_{\xi} f_{\xi \xi} + c^2 \delta^2 f_{\xi \xi \xi \xi}.
\end{equation*}
The function $u^{(1)}$ will grow linearly in $\eta = x + c t$, unless
\begin{equation*}
	2 c f_{\xi T} + c^2 f_{\xi} f_{\xi \xi} + c^2 \delta^2 f_{\xi \xi \xi \xi} = 0.
\end{equation*}
By setting $q = \frac{f_{\xi}}{6}, \tau = \frac{c T}{2}$, this equation reduces to the canonical form of the KdV equation
\begin{equation}
	q_{\tau} + 6 q q_{\xi} + \delta^2 q_{\xi \xi \xi} = 0.
	\label{4}
\end{equation}
Zabusky and Kruskal numerically studied the dynamics of the KdV equation with sinusoidal initial conditions (for small $\delta^2$, periodic boundary conditions), and discovered that the appearing solitons interact with each other elastically. They have called the emerging localised waves \textit{solitons} because of the analogy with particles. 
\begin{figure}[ht]
	\center
	\includegraphics[width = 0.5\textwidth]{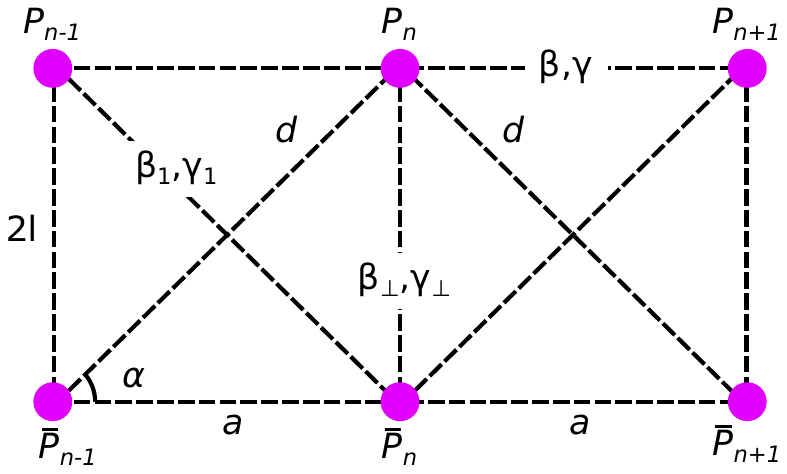}
	\caption{The dipole lattice model.}
	\label{fig:Dipole}
\end{figure} 

An extension of the FPU model in the form of the \textit{dipole lattice model} (Figure \ref{fig:Dipole})  was used in \cite{KSZ} to obtain an equation describing nonlinear longitudinal bulk strain waves in a bar.  It is convenient to view this system of coupled chains of particles as an anharmonic chain of oscillating dipoles $(P_n, \bar P_n)$ with 4 degrees of freedom: horizontal ($u_1^n$) and vertical  ($u_2^n$) displacements of the geometrical centre of a dipole, in-plane rotation of the dipole axis (to an angle $\Delta \varphi^n$), and change of a distance between the two poles of a dipole ($2 u_4^n$). This is a symmetric version of the model proposed and considered in the linear approximation in connection with the dynamics of thin films in \cite{Khusnutdinova1}. The model generalises the chain of dipoles (with the fixed distance between the poles) studied as a linear model in \cite{Askar} (see also \cite{Maugin}) and \cite{Il'yushina}, and as a nonlinear model in \cite{Khusnutdinova2}. 

The displacements of poles can be expressed in terms of the dipole coordinates as follows:
\begin{align*}
U_1^n &= u_1^n-(l+u_4^n) \sin \Delta \varphi^n, \quad
U_2^n = u_2^n+(l+u_4^n) \cos \Delta \varphi^n-l, \\
\overline{U}_1^n &= u_1^n+(l+u_4^n) \sin \Delta \varphi^n, \quad
\overline{U}_2^n = u_2^n-(l+u_4^n) \cos \Delta \varphi^n+l.
\end{align*}
Assuming that rotations are small, so that $\Delta \varphi^n=u_3^n / l \ll 1,$ we use truncated Taylor expansions in order to derive equations of motion up to quadratic terms:
\begin{equation*}
U_1^n =  u_1^n-u_3^n-\frac{u_3^n u_4^n}{l}+\dots, \quad 
U_2^n =  u_2^n+u_4^n-\frac{(u_3^n)^2}{2l}+\dots, \ \mbox{etc.}
\end{equation*}	

Then, the kinetic energy of the $n$-th dipole has the form 
\[
T_n=\frac{M}{2} \left [ (\dot{u}_1^n)^2+(\dot{u}_2^n)^2+(\dot{u}_4^n)^2
+\left(1+\frac{u_4^n}{l} \right)^2 (\dot{u}_3^n)^2 \right ],
\]
where $M = 2 m$ is the dipole mass.
The potential energy of the $n$-th dipole is defined by
pairwise interactions between neighbouring particles: 
\begin{equation}
 \Phi_n=\Phi_{n,n+1}+\Phi_{\overline{n},\overline{n+1}}+\Phi_{n,\overline{n+1}}+\Phi_{\overline{n},n+1} 
+ \Phi_{n-1,n}+\Phi_{\overline{n-1},\overline{n}}+\Phi_{n-1,\overline{n}}+\Phi_{\overline{n-1},n} +  \Phi_\perp,
\label{phi}
\end{equation}
where overlines denote particles in the second (``bottom'') row. 
Let the potential energy of interaction between any two neighbouring particles have the form 
\begin{equation}
\Phi_* (\Delta r_* )=\frac{\widetilde \beta}{2} \Delta r_* ^2+\frac{\widetilde \gamma}{3} \Delta r_*^3 + \dots,
\label{phi1}
\end{equation}
where $\Delta r_*$ is the change of a distance between the particles, and $ (\widetilde \beta, \widetilde \gamma)$ denotes one of three possible pairs of interaction constants, shown in Figure \ref{fig:Dipole}.

The change of a distance  between a pole  of the $n$-th dipole ($\widetilde P_n$) and a pole  of the $(n+1)$-th dipole ($\widetilde P_{n+1}$) is given by 
\begin{equation}
\Delta r_{\widetilde{n},\widetilde{n+1}} =
\left [(\widetilde{U}_1^{n+1}-\widetilde{U}_1^n+r_0 \cos \theta_0)^2 \right.
+
\left . (\widetilde{U}_2^{n+1}-\widetilde{U}_2^n+r_0 \sin \theta_0)^2\right ]^{1/2} - r_0,
\label{deltar}
\end{equation}
where $r_0$ is the distance between the respective pair of poles in the equilibrium configuration.
The change of a distance between the poles of the $n$-th dipole is given by $\Delta r_\perp = 2 u^n_4$.

Let us introduce the differences 
$
\Delta x=\widetilde{U}_1^{n+1}-\widetilde{U}_1^n, \
\Delta y=\widetilde{U}_2^{n+1}-\widetilde{U}_2^n,
$
and assume that these differences  are small compared to $r_0$:
$
\Delta x / r_0  \ll 1, \  \Delta y / r_0 \ll 1.
$
We can use the expansions
\begin{align*}
\Delta r_{\widetilde{n},\widetilde{n+1}} &= \Delta x \cos \theta_0+\Delta y \sin \theta_0
+
\frac{1}{2 r_0} (\Delta x \sin \theta_0-\Delta y \cos \theta_0)^2\\
&~~~~- \frac{1}{2 r_0^2} (\Delta x \cos \theta_0+\Delta y \sin \theta_0)
(\Delta x \sin \theta_0-\Delta y \cos \theta_0)^2 + ...
\end{align*}
Finally, we can derive the Euler-Lagrange equations
\[
\frac{d}{d t} \left( \frac{\partial T_n}{\partial \dot{u}_i^n} \right)-
\frac{\partial T_n}{\partial u_i^n}+\frac{\partial \Phi_n}{\partial u_i^n}=0, \quad i = \overline{1, 4}
\]
describing the nonlinear dynamics of the system, and consider the continuum approximation, as discussed before, for the FPU chain. The necessary calculations are much more tedious than that in the case of the FPU chain, but they can be performed using symbolic computations (see \cite{KSZ}).

As a result of these derivations, in \cite{KSZ} it was shown that longitudinal waves asymptotically uncouple from other degrees of freedom, which become slaved to longitudinal waves, and are described by a Boussinesq-type equation asymptotically equivalent to a ``doubly dispersive equation" (DDE) \cite{Samsonov_book, Porubov_book},
earlier derived for the long longitudinal waves in a bar of rectangular cross-section using the nonlinear elasticity approach in \cite{KS}. In dimensional variables, the DDE for a bar of rectangular cross-section $\sigma = 2a \times 2b$ has the form
\begin{equation}
	f_{tt} - c^2 f_{xx} = \frac{\beta}{2 \rho} (f^2)_{xx} + \frac{J \nu^2}{\sigma} (f_{tt} - c_1^2 f_{xx})_{xx},
	\label{DDE}
\end{equation}
where 
\begin{align*}
&c = \sqrt{E/\rho}, \quad c_1 = c / \sqrt{2 (1+\nu)}, \quad J = 4ab(a^2+b^2)/3, \\
& \beta = 3 E + 2 l (1 - 2 \nu)^3 + 4 m (1+\nu)^2 (1 - 2 \nu) + 6 n \nu^2, 
\end{align*}
and $\rho$ is the density, $E$ is the Young modulus, $\nu$ is the Poisson ratio, while $l,m,n$ are the Murnaghan moduli. Nondimensionalisation, Benjamin-Bona-Mahoney (BBM)-type regularisation \cite{BBM} of the dispersive terms and scaling bring the equation to the form
\begin{equation}
	f_{tt} - f_{xx} =  \frac{1}{2} (f^2)_{xx} + f_{ttxx}.
	\label{Be}
\end{equation}
\vspace{-2em}

\begin{figure}[ht]
	\center
	\includegraphics[width = 0.6\textwidth]{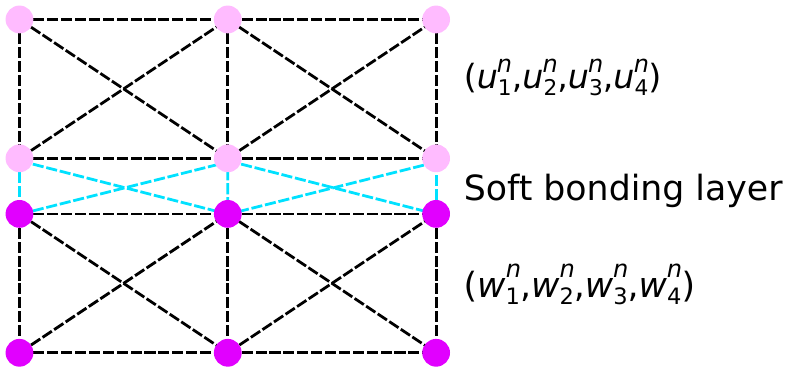}
	\caption{The layered dipole lattice model.}
	\label{fig:Lattice}
\end{figure}%
A further extension in the form of the {\it layered dipole lattice model} shown in Figure \ref{fig:Lattice} was used in \cite{KSZ} to study nonlinear waves in a bi-layer with a sufficiently soft bonding layer (e.g. some types of adhesive bonding). It was shown that longitudinal waves can be modelled by a system of coupled regularised Boussinesq (cRB) equations (presented in the scaled, nondimensional form):
\begin{align}
	f_{tt} - f_{xx} &= \frac 12 (f^2)_{xx} + f_{ttxx} - \delta (f-g)\,, \nonumber \\
	g_{tt} - c^2 g_{xx} &= \frac 12  \alpha  (g^2)_{xx} + \beta  g_{ttxx} + \gamma  (f-g)\,. \label{fg}
\end{align}
Here, $f$ and $g$ denote the  longitudinal strains in the layers, while the coefficients $c$, $\alpha$, $\beta$, $\delta$, $\gamma$ are defined by the physical and geometrical parameters of the problem (see \cite{KSZ} for the details of these derivations).

In the symmetric case ($c=\alpha=\beta=1$) the system (\ref{fg}) admits the reduction $g=f$, where $f$ satisfies the Boussinesq equation
\begin{equation}
	f_{tt} - f_{xx} = \frac{1}{2} (f^2)_{xx} + f_{ttxx}\,,
	\label{f_intro}
\end{equation}
which has particular 
solitary wave solutions:
\begin{equation}
	f = 3 (v^2 - 1) \ {\rm sech}^2 \ {\frac{\sqrt{v^2 - 1} (x - v t)}{2 \nu}}, 
	\label{soliton_intro}
\end{equation} 
where $v$ is the speed of the wave. In the cRB system of equations (\ref{fg}), when the characteristic speeds of the linear waves in the layers are close (i.e. $c$ is close to 1), this pure solitary wave solution is replaced with a radiating solitary wave \cite{KSZ,KM}, that is a solitary wave radiating a co-propagating one-sided oscillatory tail (see, for example, \cite{BGK, Bona}). The radiating solitary waves emerge due to a resonance between a soliton and a harmonic wave, which can be deduced from the analysis of the linear dispersion relation of the problem \cite{KSZ}.

\section{Scattering of pure and radiating solitary waves in bi-layers with delamination} 
\label{sec:3}

\subsection{Perfectly bonded bi-layer: scattering of a pure solitary wave}

\begin{figure}[ht]
	\center
	\includegraphics[width = 0.4\textwidth]{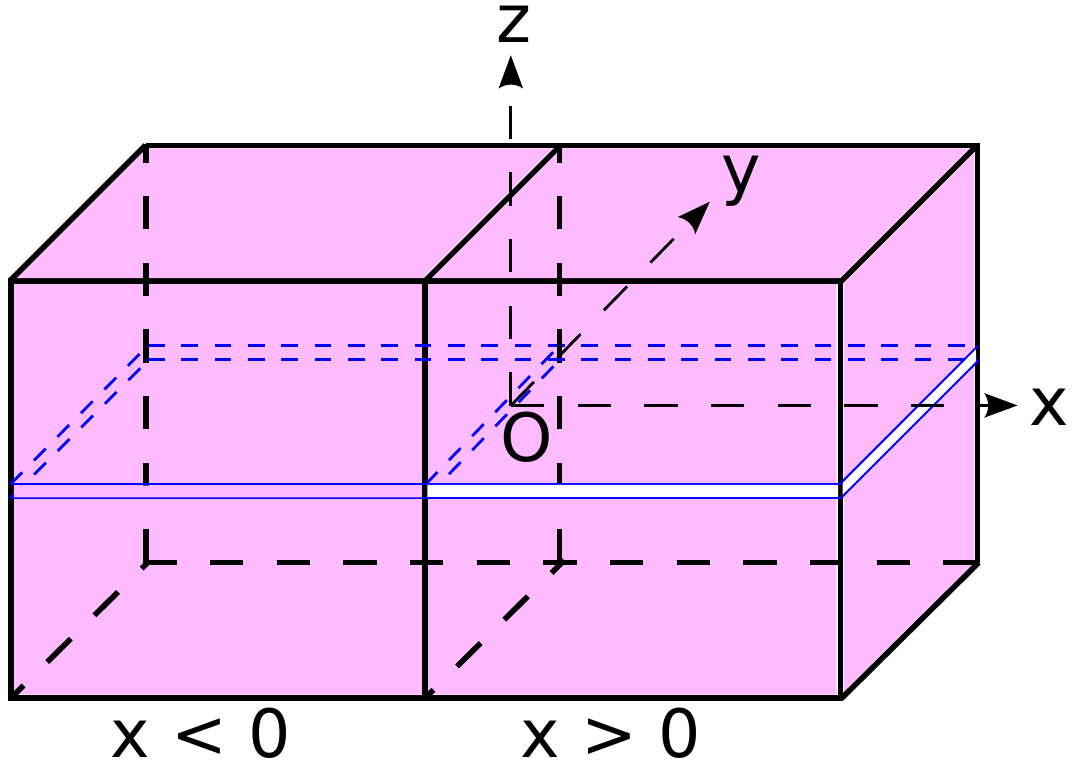}
	\caption{An example of a bi-layer with a perfect bond for $x < 0$ and complete delamination for $x > 0$.}
	\label{fig:DelamBar2S}
\end{figure}

In this section we model the scattering of a long longitudinal strain solitary wave in a perfectly bonded bi-layer with delamination at $x>0$ (see Figure \ref{fig:DelamBar2S}). The material of the layers is assumed to be the same (symmetric bar), while the material to the left and to the right of the $x=0$ cross-section can be different. 

\noindent The problem is described by the set of scaled, nondimensional equations \cite{KS}
\begin{align}
u_{tt}^{-} - u_{xx}^{-} &= \varepsilon \lsq -12 u_x^{-} u_{xx}^{-} + 2u_{ttxx}^{-} \rsq, \quad x < 0, \nonumber \\
u_{tt}^{+} - c^2 u_{xx}^{+} &= \varepsilon \lsq -12 \alpha u_x^{+} u_{xx}^{+} + 2\frac{\beta}{c^2} u_{ttxx}^{+} \rsq, \quad x > 0, \label{SB}
\end{align}
with 
associated continuity conditions
\begin{align}
u^{-}|_{x=0} &= u^{+}|_{x=0}, \label{Con1} \\
\left. u_x^{-} + \varepsilon \lsq -6 \lb u_x^{-} \rb^2  + 2u_{ttx}^{-} \rsq \right |_{x=0} &= \left. c^2 u_{x}^{+} + \varepsilon \lsq -6\alpha \lb u_{x}^{+} \rb^2 + 2\frac{\beta}{c^2} u_{ttx}^{+} \rsq \right |_{x=0}, \label{Con2}
\end{align}
and appropriate initial and boundary conditions. Here, $c, \alpha, \beta$ are constants defined by the geometrical and physical parameters of the structure, while $\varepsilon$ is the small wave amplitude parameter. The functions $u^{-}(x,t)$  and $u^{+}(x,t)$ describe displacements in the bonded and delaminated areas of the structure respectively. Condition (\ref{Con1}) is continuity of longitudinal displacement, while condition (\ref{Con2}) is the continuity of stress. In what follows we assume that $\alpha = 1$ and $\beta = \frac{n^2 + k^2}{n^2 \lb 1 + k^2 \rb}$, where $n$ represents the number of layers in the structure and $k$ is defined by the geometry of the waveguide (see \cite{KS}). The cross-section $x=0$ has width $2a$ and the height of each layer is $b$. In terms of these values $k = b/a$ and, as there are two layers in this example, $n=2$. 

In \cite{KT1} we proposed direct and semi-analytical numerical approaches to solving these two boundary-value problems matched at $x=0$. 
The semi-analytical numerical approach was based on the weakly-nonlinear solution constructed in \cite{KS}. The direct method was based on the use of finite-difference techniques \cite{KT1}.
This direct method was limited to only two sections in the structure at a time, so it could not be used for a short delamination region. The method was extended in \cite{T18} so that it can solve for multiple sections in the bar and for multiple layers.

Here we will outline the weakly-nonlinear solution derived in \cite{KS} (see also \cite{KT1}) and present the leading-order solution. To this end we look for a solution of the form
\begin{align*}
	u^{-} &= I \lb \xi_{-}, X \rb + R \lb \eta_{-}, X \rb + \varepsilon P \lb \xi_{-}, \eta_{-}, X \rb + O \lb \varepsilon^2 \rb, \\
	u^{+} &= T \lb \xi_{+}, X \rb + \varepsilon Q \lb \xi_{+}, \eta_{+}, X \rb + O \lb \varepsilon^2 \rb,
\end{align*}
where the characteristic variables are given by $\xi_{-} = x - t$, $\xi_{+} = x - ct$, $\eta_{-} = x + t$, $\eta_{+} = x + ct$, $X = \varepsilon x$. The leading-order incident, reflected and transmitted waves are described by the functions $I \lb \xi_{-}, X \rb$, $R \lb \eta_{-}, X \rb$, $T \lb \xi_{+}, X \rb$ respectively. The functions $P \lb \xi_{-}, \eta_{-}, X \rb$ and $Q \lb \xi_{+}, \eta_{+}, X \rb$ describe the higher-order corrections. Substituting this weakly-nonlinear solution into \eqref{SB} we find that the system is satisfied at leading order, while at $O \lb \varepsilon \rb$ we have
\begin{align}
	-2P_{\xi_{-} \eta_{-}} =& \lb I_{X} - 3I_{\xi_{-}}^2 + I_{\xi_{-} \xi_{-} \xi_{-}} \rb_{\xi_{-}} + \lb R_{X} - 3R_{\eta_{-}}^2 + R_{\eta_{-} \eta_{-} \eta_{-}} \rb_{\eta_{-}} \notag \\
	& - 6 \lb R I_{\xi_{-}} + I R_{\eta_{-}} \rb_{\xi_{-} \eta_{-}}. \label{Pexp}
\end{align}
To leading order the right-propagating incident wave,
$
	I = \int \tilde{I} \dd{\xi_{-}},
$
is defined by the solution of the KdV equation
\begin{equation}
	\tilde{I}_{X} - 6 \tilde{I} \tilde{I}_{\xi_{-}} + \tilde{I}_{\xi_{-} \xi_{-} \xi_{-}} = 0.
	\label{IKdV}
\end{equation}
Similarly the reflected wave,
$
	R = \int \tilde{R} \dd{\eta_{-}},
$
satisfies the KdV equation
\begin{equation}
	\tilde{R}_{X} - 6 \tilde{R} \tilde{R}_{\eta_{-}} + \tilde{R}_{\eta_{-} \eta_{-} \eta_{-}} = 0.
\label{RKdV}
\end{equation}
Expressions for the higher-order corrections can be found by substituting (\ref{IKdV}) and (\ref{RKdV}) into \eqref{Pexp} and integrating with respect to both characteristic variables, see \cite{KS} for details.

Following the same steps for the second equation in \eqref{SB}, we find that the leading-order transmitted wave,
$
	T = \int \tilde{T} \dd{\xi_{+}},
$
is described by the equation
\begin{equation}
	\tilde{T}_{X} - \frac{6}{c^2} \tilde{T} \tilde{T}_{\xi_{+}} + \frac{\beta}{c^2} \tilde{T}_{\xi_{+} \xi_{+} \xi_{+}} = 0.
\label{TKdV}
\end{equation}
To determine `initial conditions' for (\ref{RKdV}) and (\ref{TKdV}) we substitute the weakly-nonlinear solution into continuity conditions (\ref{Con1}) and (\ref{Con2}) and retain terms at leading order. Firstly, 
we differentiate \eqref{Con1} with respect to time, giving
\begin{equation*}
	u_{t}^{-}|_{x=0} = u_{t}^{+}|_{x=0},
\end{equation*}
so to leading order we have
\begin{equation}
	I_{\xi_{-}} |_{x=0} - R_{\eta_{-}} |_{x=0} = c T_{\xi_{+}} |_{x=0},
\label{ICon1}
\end{equation}
and from \eqref{Con2} we have, to leading order,
\begin{equation}
	I_{\xi_{-}} |_{x=0} + R_{\eta_{-}} |_{x=0} = c^2 T_{\xi_{+}} |_{x=0}.
\label{ICon2}
\end{equation}
Recalling that $\tilde{I} = I_{\xi_{-}}$ and similar relations for $R$ and $T$, we solve the system \eqref{ICon1}, \eqref{ICon2} for $\tilde{R}|_{x=0}$ and $\tilde{T}|_{x=0}$ in terms of $\tilde{I}|_{x=0}$ to obtain
\begin{equation}
	\tilde{R} |_{x=0} = C_{R} \tilde{I} |_{x=0}, \quad \tilde{T} |_{x=0} = C_{T} \tilde{I} |_{x=0},
	\label{IIC}
\end{equation}
where
\begin{equation}
	C_{R} = \frac{c - 1}{c + 1}, \quad C_{T} = \frac{2}{c \lb c + 1 \rb},
	\label{CRT}
\end{equation}
are the reflection and transmission coefficients respectively. This shows that, if the materials are the same in the bonded and delaminated areas ($c = 1$), then $C_{T} = 1$, $C_{R} = 0$ and there is no leading-order reflected wave.

\begin{figure}[ht]
	\center
	\includegraphics[width = 0.9\textwidth]{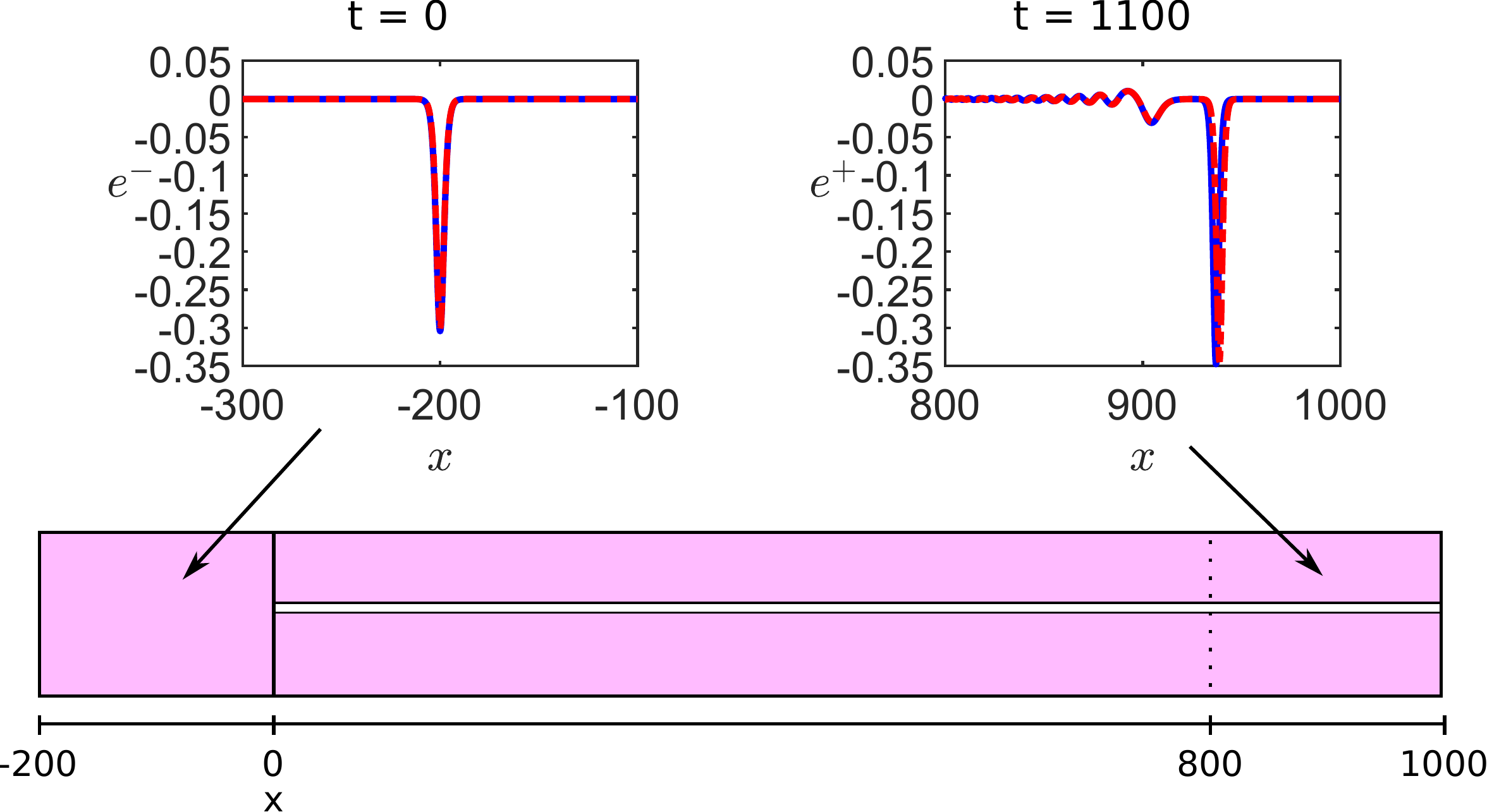}
	\caption{The solution $e^{-}$ and $e^{+}$ at the initial moment of time and at $t = 1100$, for the parameters $\alpha = 1$, $\beta = 5/8$, $c = 1$, $\varepsilon = 0.05$, with initial position $x = -200$ and initial speed $v = 1.03$, for direct numerical simulations (blue, solid line) and weakly-nonlinear solution (red, dashed line).}
	\label{fig:PB2S}
\end{figure}

A typical scenario is shown in Figure \ref{fig:PB2S} for the incident wave in the form of an exact solitary wave solution, with parameters $\alpha = 1$, $\beta = 5/8$, $c = 1$, $\varepsilon = 0.05$, with initial position $x = -200$ and initial speed $v = 1.03$. The weakly-nonlinear solution (solution of \eqref{IKdV}, \eqref{RKdV} and \eqref{TKdV}) and the results of direct numerical simulations (solution of \eqref{SB} - \eqref{Con2}) are presented for the strains $u^{-}_{x} = e^{-}$ and $u^{+}_{x} = e^{+}$ at the initial moment of time and at $t = 1100$ respectively, when the wave is propagating in the delaminated section of the bar. The results are in good agreement, with a small phase shift for the lead soliton. Soliton fission occurs in the delaminated section of the bar, with two solitons generated from a single incident soliton \cite{KS} (see also \cite{TZ, P} for the first studies of soliton fission in the context of water waves). As the material in both sections of the bar is the same, there is no leading-order reflected wave.

\begin{figure}[ht]
	\center
	\includegraphics[width = 0.9\textwidth]{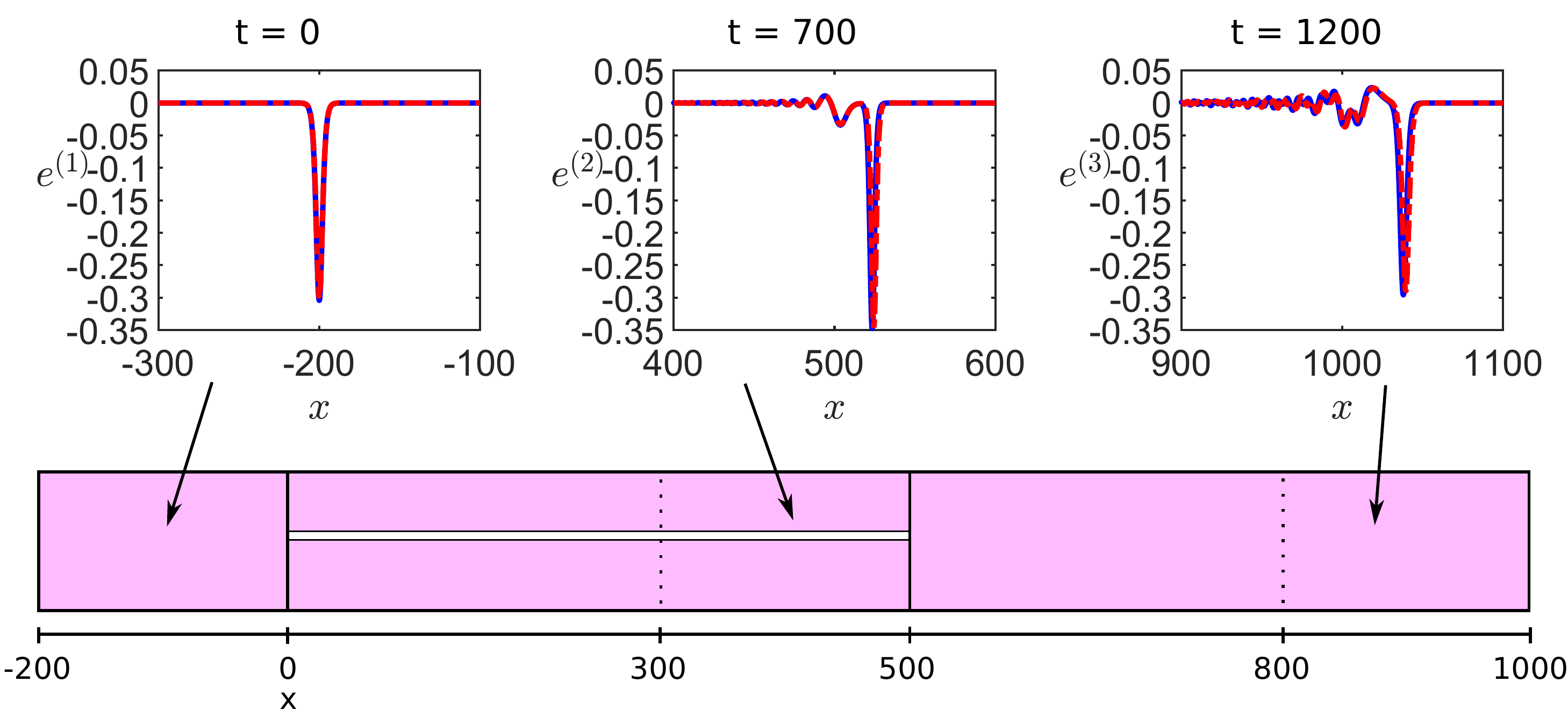}
	\caption{The solutions $e^{(1)}$, $e^{(2)}$ and $e^{(3)}$ in a bi-layer with three sections for the parameters $\alpha = 1$, $c = 1$, $\varepsilon = 0.05$ in all section and $\beta = 1$ in the bonded sections with $\beta = 5/8$ in the delaminated region. The initial position is $x = -200$ and initial speed $v = 1.03$, for direct numerical simulations (blue, solid line) and weakly-nonlinear solution (red, dashed line).}
	\label{fig:PB3S}
\end{figure}

In physical applications it is often of interest to consider the case when the delamination is finite. 
In this paper we will present a single case for a bi-layer with three sections, similarly to \cite{T18}.
We take the same parameters as for Figure \ref{fig:PB2S} and present the results in Figure \ref{fig:PB3S}, where we denote the strains in section 1 as $e^{(1)}$, and so on for sections 2 and 3. We can see that again there is good agreement between the direct numerical simulations and the semi-analytical solution constructed using the weakly-nonlinear solution. The incident soliton fissions in the delaminated region into two solitons and then, when entering the second bonded region, these solitons evolve into the solitons for the bonded region, which has different parameters, and some radiation.

An important observation in this third region is that there are now two solitons, not one, and therefore the amplitude of the lead soliton will be lower than the incident soliton. We can predict the amplitude of the solitons using the Inverse Scattering Transform (IST) as the transmitted waves in each region are described by the KdV equation.
From the IST theory  \cite{GGKM} we can determine the soliton amplitude in each region assuming that the solitons are well separated. Therefore, for a sufficiently long delaminated region we can theoretically predict the amplitude in the second bonded region. However, as the length of the delamination region is reduced, the amplitude of the soliton in the second bonded region will tend towards that of the incident soliton. This gives a quantitative measure of the delamination length \cite{T18}.


\subsection{Imperfectly bonded bi-layer: scattering of a radiating solitary wave}

\begin{figure}[ht]
	\center
	\subfigure[Homogeneous section on the left.]{\includegraphics[width=0.45\textwidth]{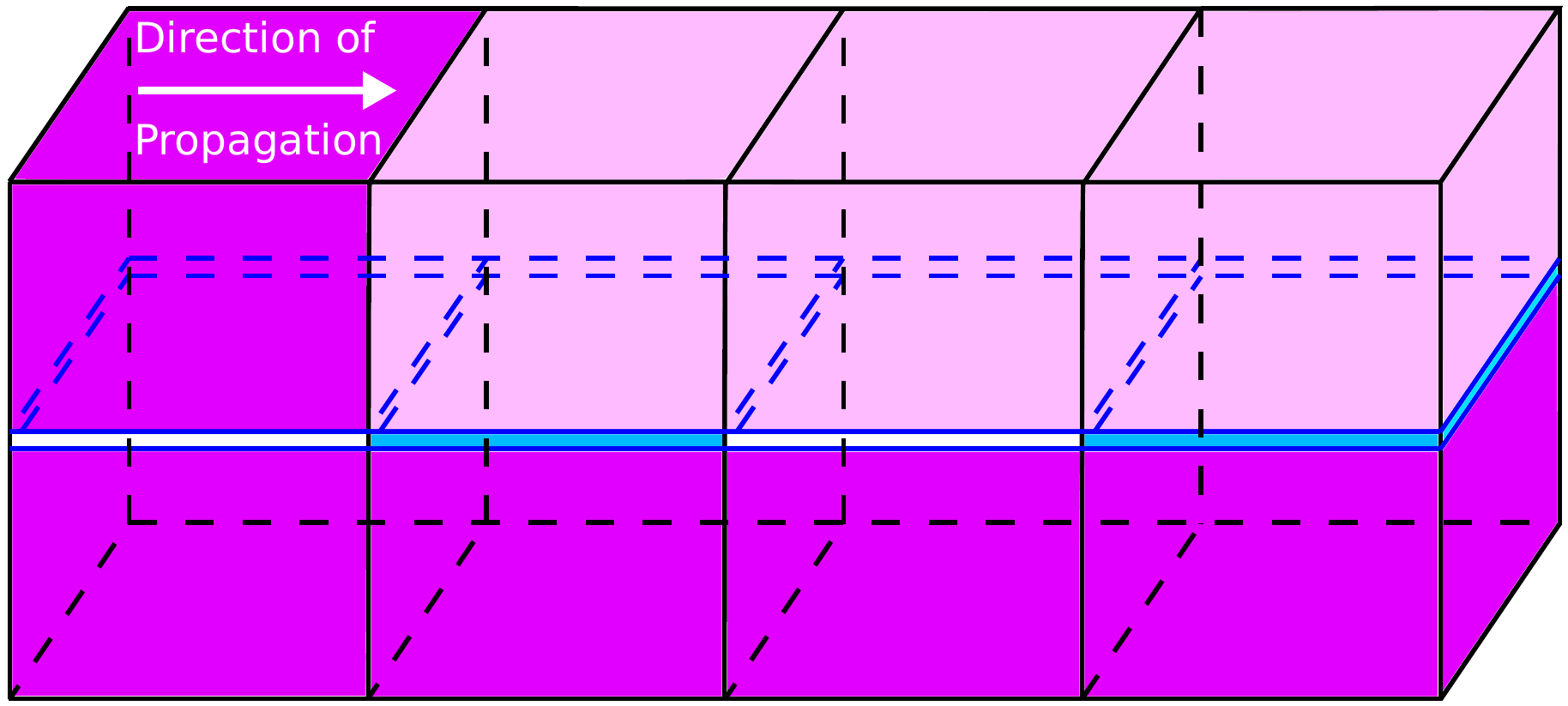}}
	\hspace{0.05\textwidth}
	\subfigure[Homogeneous section on the right.]{\includegraphics[width=0.45\textwidth]{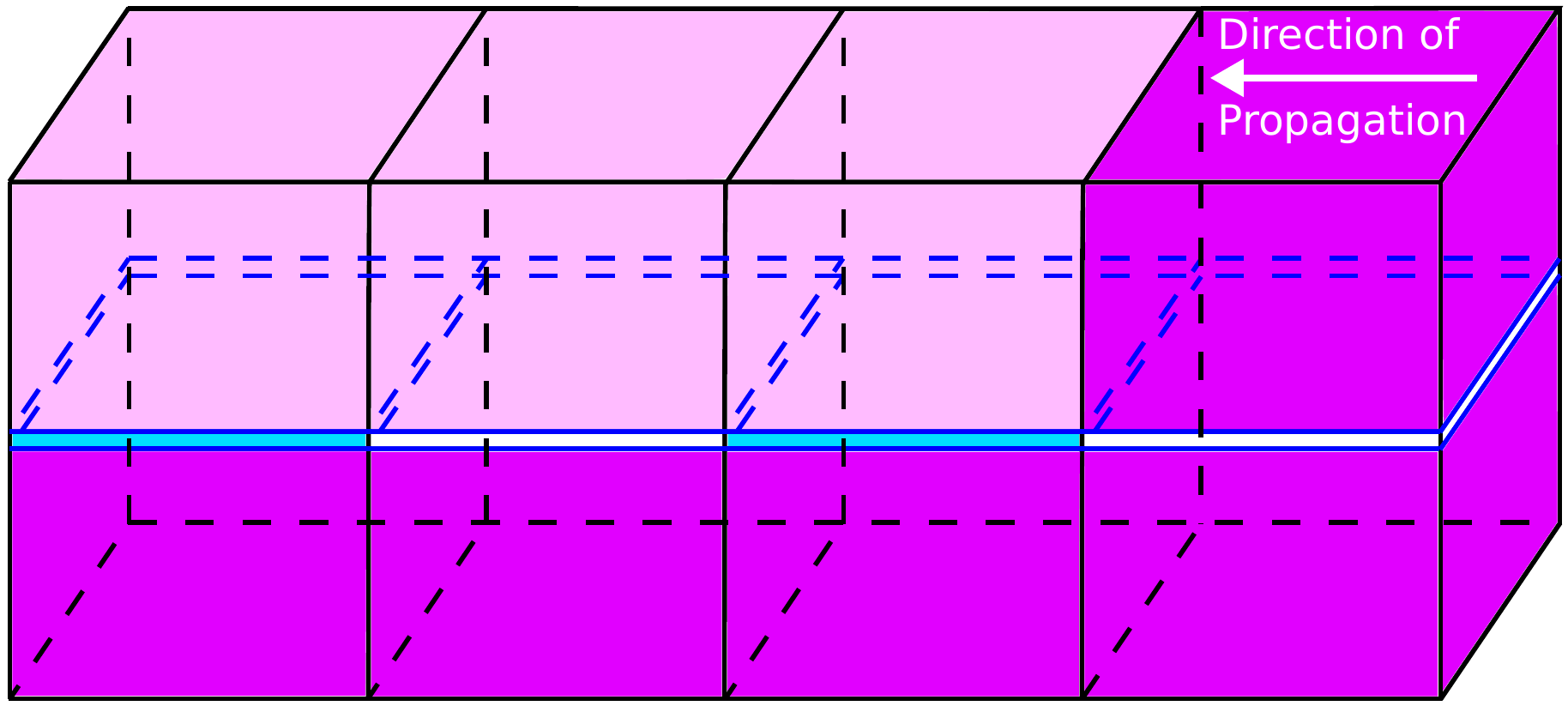}}
	\caption{An example of a bi-layer with a soft bond between the layers and a delaminated region between the bonded regions. A homogeneous section of the same material as the lower layer is attached to (a) the left-hand side of the structure, or (b) the right-hand side of the structure.}
	\label{fig:SoftBondBar}
\end{figure}

In this section we model the generation and the scattering of a long radiating solitary wave in a two-layered imperfectly bonded bi-layer with delamination (see Figure \ref{fig:SoftBondBar}). Two identical homogeneous layers (the section on the left in Figure \ref{fig:SoftBondBar}(a) and on the right in Figure \ref{fig:SoftBondBar}(b)) are ``glued"  to a two-layered structure with soft bonding between its layers, followed by a delaminated section in the middle, and another bonded section. The materials in the bi-layer are assumed to have close properties, leading to the generation of a radiating solitary wave in the bonded section.  
We model the scattering of this wave by the subsequent delaminated region, as well as the dynamics in the second bonded region. Note that the homogeneous layers can be constructed of the same material as the top layer of the bar (as was done in \cite{KT2}) but using the same material as the lower layer will give a larger amplitude, as the characteristic speed is higher which leads to a higher transmission coefficient.

The problem is described by the following sets of scaled, nondimensional equations in the respective sections of the complex waveguide \cite{KSZ, KS}:
\begin{align}
	u_{tt}^{(1)} - c^2 u_{xx}^{(1)} &= 2 \varepsilon \lsq -6 \alpha u_{x}^{(1)} u_{xx}^{(1)} + \beta u_{ttxx}^{(1)} \rsq, \nonumber \\
	w_{tt}^{(1)} - c^2 w_{xx}^{(1)} &= 2 \varepsilon \lsq -6 \alpha w_{x}^{(1)} w_{xx}^{(1)} + \beta w_{ttxx}^{(1)} \rsq, \label{syshomog}
\end{align} 
for the section with two homogeneous layers,
\begin{align}
u_{tt}^{(2,4)} - u_{xx}^{(2,4)} &= 2\varepsilon \lsq -6 u_{x}^{(2,4)} u_{xx}^{(2,4)} + u_{ttxx}^{(2,4)} -~ \delta \lb u^{(2,4)} - w^{(2,4)} \rb \rsq, \nonumber \\
w_{tt}^{(2,4)} - c^2 w_{xx}^{(2,4)} &= 2 \varepsilon \lsq -6 \alpha w_{x}^{(2,4)} w_{xx}^{(2,4)} + \beta w_{ttxx}^{(2,4)}  +~ \gamma \lb u^{(2,4)} - w^{(2,4)} \rb \rsq, \label{syscoup} 
\end{align}
for the two bonded regions, and
\begin{align}
u_{tt}^{(3)} - u_{xx}^{(3)} &= 2 \varepsilon \lsq -6 u_{x}^{(3)} u_{xx}^{(3)} + u_{ttxx}^{(3)} \rsq, \nonumber \\
w_{tt}^{(3)} - c^2 w_{xx}^{(3)} &= 2 \varepsilon \lsq -6 \alpha w_{x}^{(3)} w_{xx}^{(3)} + \beta w_{ttxx}^{(3)} \rsq, \label{sysdelam}
\end{align}
for the delaminated region.
Here the functions $u^{(i)}(x,t)$ and $w^{(i)}(x,t)$ describe longitudinal displacements in the ``top" and ``bottom" layers of the four sections of the waveguide, respectively. The values of the constants $\alpha$, $\beta$ and $c$ depend on the physical and geometrical properties of the waveguide, while the constants $\delta$ and $\gamma$ depend on the properties of the soft bonding layer, and $\varepsilon$ is a small amplitude parameter (see \cite{KSZ, KT2}).

These equations are complemented with continuity conditions for the longitudinal displacements and stresses at the interfaces between the sections, similarly to (\ref{Con1}), (\ref{Con2}), as well as the relevant initial and boundary conditions  (see \cite{KT1, KT2} for details).

Again, the direct numerical modelling of this problem is difficult and expensive because one needs to solve several boundary value problems linked to each other via matching conditions at the boundaries. Therefore, we developed an alternative semi-analytical approach based upon the use of several matched asymptotic multiple-scales expansions and averaging with respect to the fast space variable at a constant value of one or another characteristic variable \cite{KT1, KT2}. We present an outline of the method here, omitting the matching at the interface between sections of the bi-layer which can be found in \cite{KT1, KT2}.

To determine the weakly-nonlinear solution to this system of equations, we firstly differentiate the governing equations with respect to $x$ so that we can use a space averaging procedure, as the strain waves are localised. We denote $f^{(i)} = u_x^{(i)}$ and $g^{(i)} = w_x^{(i)}$ to obtain the equations `in strains'
\begin{align}
f_{tt}^{(1)} - c^2 f_{xx}^{(1)} &= 2 \varepsilon \lsq -3 \alpha \lb f^{(1)} \rb^2 + \beta f_{tt}^{(1)} \rsq _{xx}, \notag \\
g_{tt}^{(1)} - c^2 g_{xx}^{(1)} &= 2 \varepsilon \lsq -3 \alpha \lb g^{(1)} \rb^2 + \beta g_{tt}^{(1)} \rsq _{xx}, \label{syshomogdiff} 
\end{align}
for $x < x_{a}$,
\begin{align}
f_{tt}^{(2,4)} - f_{xx}^{(2,4)} &= 2 \varepsilon \lsq -3 \lb f^{(2,4)} \rb^2 + f_{tt}^{(2,4)} \rsq_{xx} - 2 \varepsilon \delta \lb f^{(2,4)} - g^{(2,4)} \rb, \notag \\
g_{tt}^{(2,4)} - c^2 g_{xx}^{(2,4)} &= 2 \varepsilon \lsq -3 \alpha \lb g^{(2,4)} \rb^2 + \beta g_{tt}^{(2,4)} \rsq_{xx} + 2 \varepsilon \gamma \lb f^{(2,4)} - g^{(2,4)} \rb, \label{syscoupdiff} 
\end{align}
for $x_{a} < x < x_{b}$ and $x > x_{c}$ (bonded regions) and 
\begin{align}
f_{tt}^{(3)} - f_{xx}^{(3)} &= 2 \varepsilon \lsq -3 \lb f^{(3)} \rb^2 + f_{tt}^{(3)} \rsq _{xx}, \notag \\
g_{tt}^{(3)} - c^2 g_{xx}^{(3)} &= 2 \varepsilon \lsq -3 \alpha \lb g^{(3)} \rb^2 + \beta g_{tt}^{(3)} \rsq  _{xx} \label{sysdelamdiff}
\end{align}
for $x_{b} < x < x_{c}$ (delaminated region). 
 
We assume that all functions present in our expansions and their derivatives are bounded and sufficiently rapidly decaying at infinity. In the first region, the equation is identical in both homogeneous layers and therefore we assume the same incident wave in both, and consider asymptotic multiple-scales expansions of the type
\begin{align*}
f^{(1)} = I \lb \nu, X \rb + R^{(1)} \lb \zeta, X \rb + \varepsilon P^{(1)} \lb \nu, \zeta, X \rb + O \lb \varepsilon^2 \rb, \\
g^{(1)} = I \lb \nu, X \rb + G^{(1)} \lb \zeta, X \rb + \varepsilon Q^{(1)} \lb \nu, \zeta, X \rb + O \lb \varepsilon^2 \rb, 
\end{align*}
where the characteristic variables are given by $\nu = x - ct$, $\zeta = x + ct$,  and  the slow variable $X = \varepsilon x$. Here, the functions $I$ and $R^{(1)}$, $G^{(1)}$ represent the leading-order incident and reflected waves respectively and $P^{(1)}, Q^{(1)}$ are the higher-order corrections. Substituting the asymptotic expansion into the first equation in \eqref{syshomogdiff} and applying the averaging with respect to the fast space variable we obtain the equations 
\begin{align}
I_{X} - 6 \frac{\alpha}{c^2} I I_{\nu} + \beta I_{\nu \nu \nu} &= 0, \label{Ieq} \\
R^{(1)}_{X} - 6 \frac{\alpha}{c^2} R^{(1)} R^{(1)}_{\zeta} + \beta R^{(1)}_{\zeta \zeta \zeta} &= 0. \label{Req}
\end{align}
Similarly for the second layer we obtain	
\begin{equation}
G^{(1)}_{X} - 6 \frac{\alpha}{c^2} G^{(1)} G^{(1)}_{\zeta} + \beta G^{(1)}_{\zeta \zeta \zeta} = 0, \label{Geq}
\end{equation}
in addition to \eqref{Ieq}.

For the second section of the bar we expect radiating solitary waves to develop if the layers have close properties. We seek a weakly-nonlinear solution of the form 
\begin{align*}
f^{(2)} &= T^{(2)} \lb \xi, X \rb + R^{(2)} \lb \eta, X \rb + \varepsilon P^{(2)} \lb \xi, \eta, X \rb + O \lb \varepsilon^2 \rb,  \\
g^{(2)} &= S^{(2)} \lb \xi, X \rb + G^{(2)} \lb \eta, X \rb + \varepsilon Q^{(2)} \lb \xi, \eta, X \rb + O \lb \varepsilon^2 \rb.
\end{align*}
The characteristic variables are $\xi = x - t$, $\eta = x + t$ and $X$ is the same as before, $T^{(2)}$ and $S^{(2)}$ represent the transmitted waves in the second section of the bar, where $T$ is for the top layer and $S$ is for the bottom layer. Similarly, $R^{(2)}$ and $G^{(2)}$ are the reflected waves, and the higher-order corrections in this section are given by $P^{(2)}$ and $Q^{(2)}$, for the top and bottom layers respectively. We substitute this weakly-nonlinear solution into (\ref{syscoupdiff}) and apply the averaging to obtain the system of equations
\begin{align}
& \lb T^{(2)}_{X} - 6 T^{(2)} T^{(2)}_{\xi} + T^{(2)}_{\xi \xi \xi} \rb_{\xi} = \delta \lb T^{(2)} - S^{(2)} \rb,  \label{T2eq} \\
& \lb S^{(2)}_{X} + \frac{c^2 - 1}{2 \varepsilon} S^{(2)}_{\xi} - 6 \alpha S^{(2)} S^{(2)}_{\xi} + \beta S^{(2)}_{\xi \xi \xi} \rb_{\xi} = \gamma \lb S^{(2)} - T^{(2)} \rb.
\label{S2eq}
\end{align}
This is a system of {\it coupled Ostrovsky equations} (note that the Ostrovsky equation was initially derived to describe long surface and internal waves in a rotating ocean \cite{O, review}), which appear naturally in the description of nonlinear waves in layered waveguides, both solid and fluid \cite{KM, AGK}.

Similarly, for the reflected waves in this region, we obtain 
\begin{align}
&\lb R^{(2)}_{X} - 6 R^{(2)} R^{(2)}_{\eta} + R^{(2)}_{\eta \eta \eta} \rb_{\eta} = \delta \lb R^{(2)} - G^{(2)} \rb, \label{R2eq} \\
& \lb G^{(2)}_{X} + \frac{c^2 - 1}{2 \varepsilon} G^{(2)}_{\eta} - 6 \alpha G^{(2)} G^{(2)}_{\eta} + \beta G^{(2)}_{\eta \eta \eta} \rb_{\eta} = \gamma \lb G^{(2)} - R^{(2)} \rb.
\label{G2eq}
\end{align}
Therefore, to leading order, the transmitted and reflected waves are described by two systems of coupled Ostrovsky equations.

We now consider the delaminated region and look for a weakly-nonlinear solution to \eqref{sysdelamdiff} of the form
\begin{align*}
f^{(3)} &= T^{(3)} \lb \xi, X \rb + R^{(3)} \lb \eta, X \rb + \varepsilon P^{(2)} \lb \xi, \eta, X \rb + O \lb \varepsilon^2 \rb,  \\
g^{(3)} &= S^{(3)} \lb \nu, X \rb + G^{(3)} \lb \zeta, X \rb + \varepsilon Q^{(2)} \lb \nu, \zeta, X \rb + O \lb \varepsilon^2 \rb,
\end{align*}
where the characteristic variables are the same as in previous sections. Substituting this into system \eqref{sysdelamdiff} and averaging with respect to the fast space variable we obtain the equations
\begin{align}
T^{(3)}_{X} - 6 T^{(3)} T^{(3)}_{\xi} + T^{(3)}_{\xi \xi \xi} &= 0, \label{T3eq} \\
S^{(3)}_{X} - 6 \frac{\alpha}{c^2} S^{(3)} S^{(3)}_{\nu} + \beta S^{(3)}_{\nu \nu \nu} &= 0, \label{S3eq}
\end{align}
describing transmitted waves, and the reflected waves are governed by the equations
\begin{align}
R^{(3)}_{X} - 6 R^{(3)} R^{(3)}_{\eta} + R^{(3)}_{\eta \eta \eta} &= 0, \label{R3eq} \\
G^{(3)}_{X} - 6 \frac{\alpha}{c^2} G^{(3)} G^{(3)}_{\zeta} + \beta G^{(3)}_{\zeta \zeta \zeta} &= 0. \label{G3eq}
\end{align}
Finally, in the fourth region, we can make use of the same weakly-nonlinear solution that was used in the second region and we obtain the same equations, but only for the functions describing transmitted waves in this region (no boundary to generate reflected waves). Therefore, with the weakly-nonlinear solution
\begin{align*}
f^{(4)} &= T^{(4)} \lb \xi, X \rb + \varepsilon P^{(4)} \lb \xi, \eta, X \rb + O \lb \varepsilon^2 \rb,  \\
g^{(4)} &= S^{(4)} \lb \xi, X \rb + \varepsilon Q^{(4)} \lb \xi, \eta, X \rb + O \lb \varepsilon^2 \rb,
\end{align*}
the transmitted waves in this region are described by the coupled Ostrovsky equations
\begin{align}
& \lb T^{(4)}_{X} - 6 T^{(4)} T^{(2)}_{\xi} + T^{(4)}_{\xi \xi \xi} \rb_{\xi} = \delta \lb T^{(4)} - S^{(4)} \rb,  \label{T4eq} \\
& \lb S^{(4)}_{X} + \frac{c^2 - 1}{2 \varepsilon} S^{(4)}_{\xi} - 6 \alpha S^{(4)} S^{(4)}_{\xi} + \beta S^{(4)}_{\xi \xi \xi} \rb_{\xi} = \gamma \lb S^{(4)} - T^{(4)} \rb.
\label{S4eq}
\end{align}
In order to find `initial conditions' for the derived equations, we collect the expressions for the weakly-nonlinear solutions and substitute them into the continuity conditions and retain terms at leading order, as was done for the perfectly bonded bi-layer. See \cite{KT1} for a detailed explanation of this procedure.

\begin{figure}[ht]
	\center
	\includegraphics[width = 1.0\textwidth]{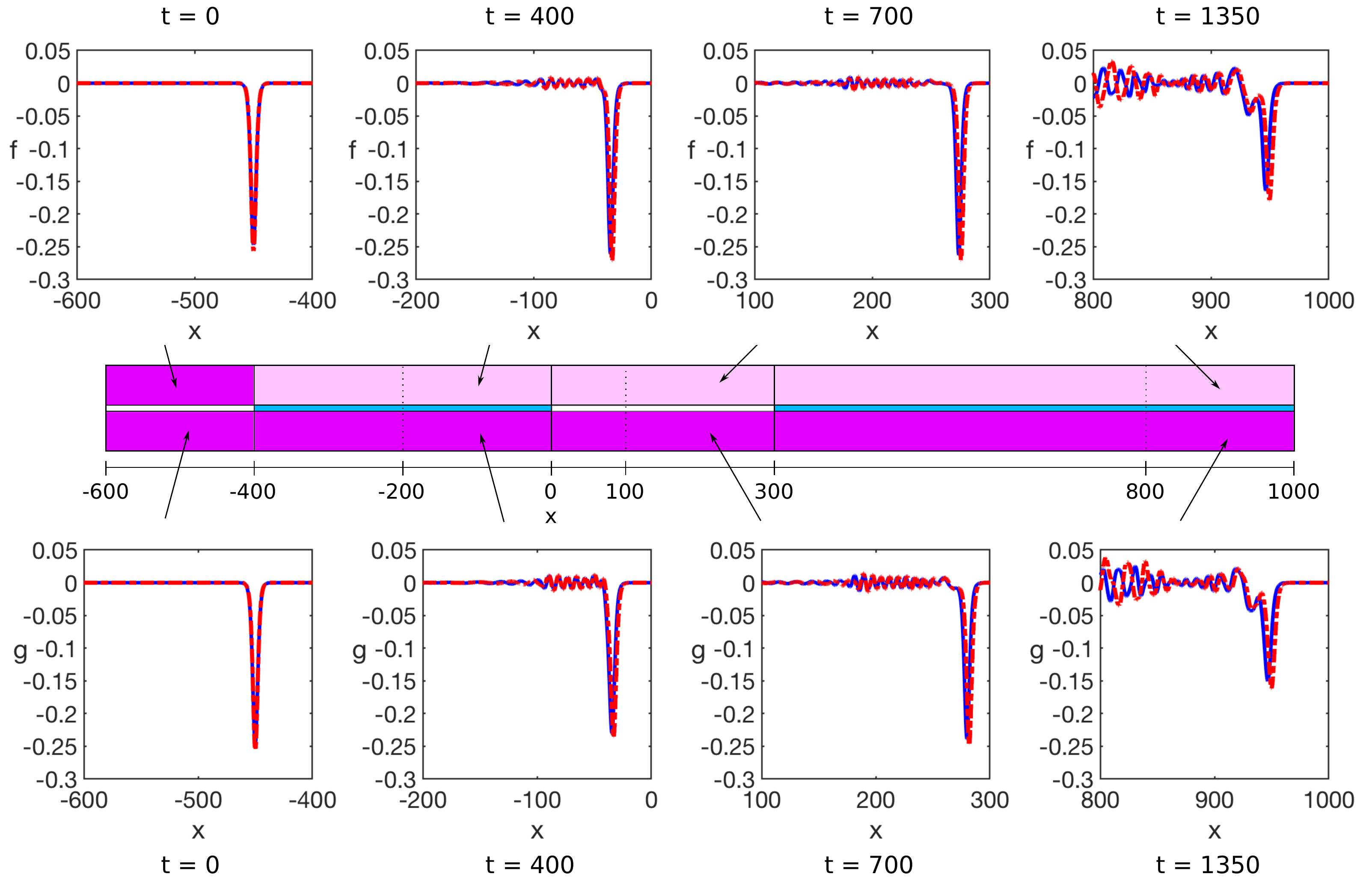}
	\caption{The solutions $f$ (top row) and $g$  (bottom row) in each section of the bi-layer, for the parameters $\alpha = \beta = 1.05$, $c = 1.025$, $\delta = \gamma = 0.8$ and $\varepsilon = 0.05$, with initial position $x = -450$ and initial speed $v = 1.05$, for direct numerical simulations (blue, solid line) and semi-analytical method (red, dashed line). Two homogeneous layers, of the same material as the lower layer, are on the left and the waves propagate to the right.}
	\label{fig:CDCM2L}
\end{figure}

A typical scenario is shown in Figure \ref{fig:CDCM2L} for the incident wave in the form of an exact solitary wave solution, with parameters $\alpha = \beta = 1.05$, $c = 1.025$, $\delta = \gamma = 0.8$ and $\varepsilon = 0.05$, with initial position $x = -450$ and initial speed $v = 1.05$. Two homogeneous layers, of the same material as the lower layer, are attached to the left of the bar. We see the generation of a radiating solitary wave in the bonded section of the bar, the separation of the solitary wave from its radiating tail in the delaminated section, and the re-coupling of the waves in the second bonded region. 

\begin{figure}[ht]
	\center
	\includegraphics[width = 1.0\textwidth]{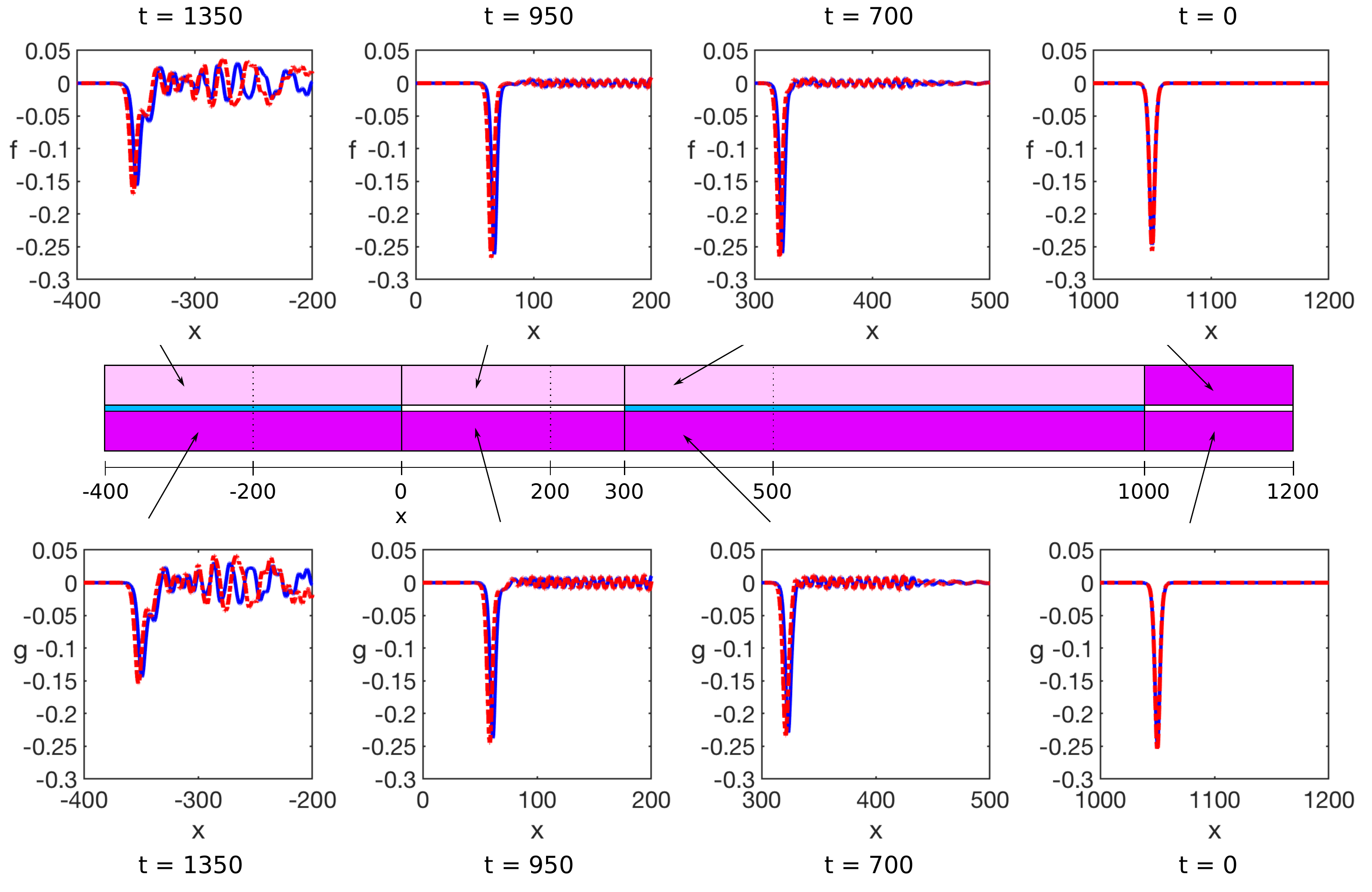}
	\caption{The solutions $f$ (top row) and $g$  (bottom row) in each section of the bi-layer, for the parameters $\alpha = \beta = 1.05$, $c = 1.025$, $\delta = \gamma = 0.8$ and $\varepsilon = 0.05$, with initial position $x = 1050$ and initial speed $v = 1.05$, for direct numerical simulations (blue, solid line) and semi-analytical method (red, dashed line). Two homogeneous layers, of the same material as the lower layer, are on the right and the waves propagate to the left.}
	\label{fig:CDCM2R}
\end{figure}

A similar numerical experiment for the same parameters, with the homogeneous section on the right-hand side of the bi-layer, is shown in Figure \ref{fig:CDCM2R}. The results in this case are qualitatively similar, with a different length of radiating tail due to the change in the length of the relevant bonded section. We note that if the radiation wave packet in the second bonded region is closer to the leading wave when sending the waves from the right, then the delamination is closer to the left-hand side of the structure, and vice versa.


\section{Initial-value problem}

When the characteristic speeds in the layers significantly differ,  the dynamical behaviour is different \cite{KM}. To simplify the problem, let us assume that the material of the lower layer has much greater density (greater inertia), and consider the reduction $w=0$. Then, the dynamics of the top layer is described (after appropriate scalings) by the {\it Boussinesq-Klein-Gordon} (BKG) equation \cite{KT3}
\begin{equation*}
u_{tt} - c^2 u_{xx} = \varepsilon \lsq \frac{\alpha}{2} \lb u^2 \rb_{xx} + \beta u_{ttxx} - \gamma u \rsq.
\end{equation*}
This case is similar to a Toda lattice on an elastic substrate \cite{YK}. Such an equation has also arisen in the context of oceanic waves in a rotating ocean \cite{G}. While the accuracy of such single Boussinesq-type equations does not exceed the accuracy of the corresponding uni-directional models  in the water-wave context, they are valid two-directional models for various solid waveguides (see, for example,  \cite{Samsonov_book,Porubov_book, EST, PT, GKS} and references therein).

The Ostrovsky equation \cite{O}, written in the general form 
\begin{equation*}
	\lb \eta_t + \nu \eta \eta_x + \mu \eta_{xxx} \rb_x = \lambda \eta
\end{equation*}
with some constant coefficients $\nu, \mu$ and $\lambda$, implies that for any regular periodic solution on a finite interval (including the case of localised solutions on a large interval)  the mean is zero for any $t\ge 0$:
\begin{equation*}
\frac{1}{2L} \int_{-L}^L \eta \dd{x} = 0.	 
\end{equation*}
However, the original BKG equation  does not impose similar restrictions on its solutions or initial conditions (apparent {\it zero-mass contradiction}). 

This issue has been settled on the infinite line by considering a regularised Ostrovsky equation \cite{G99}. Similar regularisation was  used for the Kadomtsev-Petviashvili equation \cite{AW, BPP}, and the physical motivation has been discussed in \cite{GM}. It was shown that the mass rapidly adjusts within the {\it temporal boundary layer}. The main wave has zero mass, and the non-zero mass is transported to a large distance in the opposite direction to the propagation of this wave \cite{GH}.

In this section we outline the construction of the solution which bypasses this difficulty in the periodic case, following \cite{KMP, KT3}.  We consider the initial-value problem on the interval $ [-L, L]$:
\begin{align}
	u_{tt} - c^2 u_{xx} &= \varepsilon \lsq \frac{\alpha}{2} \lb u^2 \rb_{xx} + \beta u_{ttxx} - \gamma u \rsq, \label{BousOstOld1} \\
	u |_{t=0} &= F(x), \quad u_t |_{t=0} = V(x), \label{BousOstIC1}
\end{align}
where $F$ and $V$ are sufficiently smooth $(2L)$-periodic functions, and both functions $F(x)$ and $V(x)$ may have non-zero mean values
\begin{equation}
F_{0} = \frac{1}{2L} \int_{-L}^{L} F(x) \dd{x}  \eqtext{and} V_{0} = \frac{1}{2L} \int_{-L}^{L} V(x) \dd{x}.
\label{MeanValIC1}
\end{equation}
The mean value of $u$ is calculated as
\begin{equation}
	\langle u \rangle (t) := \frac{1}{2L} \int_{-L}^{L} u(x,t) \dd{x} = F_{0} \cos{\lb \sqrt{\varepsilon \gamma} t \rb} + V_{0} \frac{\sin{\lb \sqrt{\varepsilon \gamma} t \rb}}{\sqrt{\varepsilon \gamma}}.
	\label{MeanVal1}
\end{equation}
The initial-value problem for the deviation from the oscillating mean value $\tilde{u} = u - \langle u \rangle (t)$ is given by
\begin{align}
	\tilde u_{tt} - c^2 \tilde u_{xx} = \varepsilon &\lsq \alpha \lb F_0 \cos{\lb \omega t  \rb}  + \frac{1}{\sqrt{\varepsilon}} \frac{V_0}{\sqrt{\gamma}} \sin {\lb \omega t \rb} \rb \tilde u_{xx} \right. \notag \\
	&\left. ~+ \frac{\alpha}{2} \lb \tilde u^2 \rb_{xx} + \beta \tilde u_{ttxx} - \gamma \tilde u \rsq,
	\label{BousOstEq1}
\end{align}
and
\begin{equation}
\tilde u |_{t=0} = F(x) - F_0, \quad \tilde u_t |_{t=0} = V(x) - V_0,
\label{BousOstICnew1}
\end{equation}
where $\omega = \sqrt{\gamma \varepsilon}$.

We look for a weakly-nonlinear solution of the form
\begin{align}
	\tilde u \lb x, t \rb &= f^{+} \lb \xi_{+}, \tau, T \rb +  f^{-} \lb \xi_{-}, \tau, T \rb + \sqrt{\varepsilon} P \lb \xi_{-}, \xi_{+}, \tau, T \rb + O \lb \varepsilon \rb,
	\label{WNLSol1}
\end{align}
where 
\begin{equation*}
\xi_{\pm} = x \pm c t, \quad \tau = \sqrt{\varepsilon} t, \quad T = \varepsilon t.
\end{equation*}
Here we aim to construct the solution up to and including $O \lb \sqrt{\varepsilon} \rb$ terms. 

The first non-trivial equation appears at $O \lb \sqrt{\varepsilon} \rb$:
\begin{equation}
- 4 c^2 P_{\xi_- \xi_+} = 2 c f^-_{\xi_- \tau} - 2 c f^+_{\xi_+ \tau} + \frac{\alpha V_0}{\sqrt{\gamma}} \sin \lb \sqrt{\gamma} \tau \rb \lb f^-_{\xi_- \xi_-} + f^+_{\xi_+ \xi_+} \rb.
\label{Peq}
\end{equation}
Averaging with respect to $x$ at constant $\xi_-$ or $\xi_+$ yields the equations
\begin{equation}
2 c f^-_{\xi_- \tau} + \frac{\alpha V_0}{\sqrt{\gamma}} \sin \lb \sqrt{\gamma} \tau \rb f^-_{\xi_- \xi_-} = 0,
\label{f-}
\end{equation}
and
\begin{equation}
2 c f^+_{\xi_+ \tau} - \frac{\alpha V_0}{\sqrt{\gamma}} \sin \lb \sqrt{\gamma} \tau \rb f^+_{\xi_+ \xi_+} = 0.
\label{f+}
\end{equation}
The equations are integrated using the method of characteristics, yielding
\begin{equation}
f^- = f^- \lb \xi_- + \frac{\alpha V_0}{2 c \gamma} \cos \lb \sqrt{\gamma} \tau \rb, T\rb, \quad f^+ = f^+ \lb \xi_+ - \frac{\alpha V_0}{2 c \gamma} \cos \lb \sqrt{\gamma} \tau \rb, T\rb.
\label{ff}
\end{equation}
The formulae (\ref{ff}) motivate the change of variables
\begin{equation}
\tilde \xi_- = \xi_- + \frac{\alpha V_0}{2 c \gamma} \cos \lb \sqrt{\gamma} \tau \rb, \quad \tilde \xi_+ = \xi_+ - \frac{\alpha V_0}{2 c \gamma} \cos \lb \sqrt{\gamma} \tau \rb,
\label{change}
\end{equation}
instead of $\xi_-$ and $\xi_+$, and we can now rewrite the equation for $P$ as $P_{\tilde \xi_- \tilde \xi_+} = 0$, which gives
\begin{equation}
P = g^-\lb \tilde \xi_-, \tau, T \rb + g^+ \lb \tilde \xi_+, \tau, T \rb.
\label{P}
\end{equation}
At $O \lb \varepsilon \rb$, using the averaging, we obtain
\begin{align}
g^{\pm} _{\tilde \xi_{\pm}} &= - \frac{\alpha V_0}{4 c^2 \sqrt{\gamma}} \sin \lb \sqrt{\gamma} \tau \rb f^{\pm}_{\tilde \xi_{\pm} } \pm \frac{1}{2 c} A^{\pm}\lb \tilde \xi_{\pm}, T \rb \tau \nonumber \\
&~~~~\mp \lsq  \frac{\alpha^2 V_0^2}{16 c^3 \gamma} \lb \tau - \frac{\sin \lb 2 \sqrt{\gamma} \tau \rb}{2 \sqrt{\gamma}} \rb - \frac{\alpha F_0}{2 c \sqrt{\gamma}} \sin \lb \sqrt{\gamma} \tau \rb \rsq f^{\pm}_{\tilde \xi_{\pm} \tilde \xi_{\pm}}, 
\label{g-}
\end{align}
where 
\begin{equation}
A^{\pm}  \lb  \tilde \xi_{\pm}, T \rb = \lb \mp 2 c f_{T}^{\pm} + \alpha f^{\pm} f_{\tilde \xi_{\pm}}^{\pm} + \beta c^2 f_{\tilde \xi_{\pm} \tilde \xi_{\pm} \tilde \xi_{\pm}}^{\pm} \rb_{\tilde \xi_{\pm}} - \gamma f^{\pm}.
\end{equation}
Here, we omitted the homogeneous parts of the solutions for $g^{\pm} _{\tilde \xi_{\pm}}$. They can be shown to be equal to zero (see \cite{KT3}). To avoid secular terms we require
\begin{equation}
\lb \mp 2 c f_{T}^{\pm} - \frac{\alpha^2 V_0^2}{8 c^2 \gamma}  f^{\pm}_{\tilde \xi_{\pm}}  + \alpha f^{\pm} f_{\tilde \xi_{\pm}}^{\pm} + \beta c^2 f_{\tilde \xi_{\pm} \tilde \xi_{\pm} \tilde \xi_{\pm}}^{\pm} \rb_{\tilde \xi_{\pm}} - \gamma f^{\pm} = 0.
\label{Ost1}
\end{equation}
Thus, we obtain two Ostrovsky equations for the left- and right-propagating waves. The equations (\ref{Ost1}) can be reduced to the standard form of the Ostrovsky equations by the change of variables
$$
\hat \xi_{\pm} = \tilde \xi_{\pm} \mp \frac{\alpha^2 V_0^2}{16 c^3 \gamma} T.
$$
The expansion (\ref{WNLSol1}) is also substituted into the initial conditions (\ref{BousOstICnew1}), which we satisfy at the respective orders of the small parameter.

The solution of the Cauchy problem (\ref{BousOstOld1}), (\ref{BousOstIC1}) for the original variable $u(x, t)$ up to and including $O \lb \sqrt{\varepsilon} \rb$ terms has the form
\begin{align}
u(x, t) &= V_{0} \frac{\sin{\lb \sqrt{\gamma} \tau \rb}}{\sqrt{\varepsilon \gamma}} + F_{0} \cos{\lb \sqrt{\gamma} \tau \rb}  \nonumber \\ 
&~~~~+ f^-  + f^+  + \sqrt{\varepsilon} \lsq -\frac{\alpha V_0}{4 c^2 \sqrt{\gamma}} \sin \sqrt{\gamma} \tau \lb f^- + f^+ \rb  \right . \nonumber \\
&~~~~\left . - \frac{\alpha }{2 c \sqrt{\gamma}} \lb F_0 \sin \lb \sqrt{\gamma} \tau \rb + \frac{\alpha V_0^2}{16 c^2 \gamma} \sin \lb 2 \sqrt{\gamma} \tau \rb \rb \lb f^-_{\tilde \xi_-} - f^+_{\tilde \xi_+} \rb  \rsq + O(\varepsilon),
\label{WNLFV}
\end{align}
where the functions $f^{\pm} \lb \tilde \xi_{\pm}, T \rb$ are solutions of the Ostrovsky equations (\ref{Ost1}), which should be solved subject to the initial conditions
\begin{equation}
f^{\pm}|_{T=0} = \frac{1}{2c} \lb c [F \lb \tilde \xi_{\pm} \rb - F_0] \pm \int_{-L}^{\tilde \xi_{\pm}} (V(\sigma) - V_0) \dd{\sigma} \rb.
\end{equation}
Here $\displaystyle \tilde \xi_{\pm} = \xi_{\pm} \mp \frac{\alpha V_0}{2 c \gamma} \cos \lb \sqrt{\gamma} \tau \rb$ (nonlinear characteristic variables when $V_0 \ne 0$).

\begin{figure}[!htbp]
	\centering
	\includegraphics[width=0.8\textwidth]{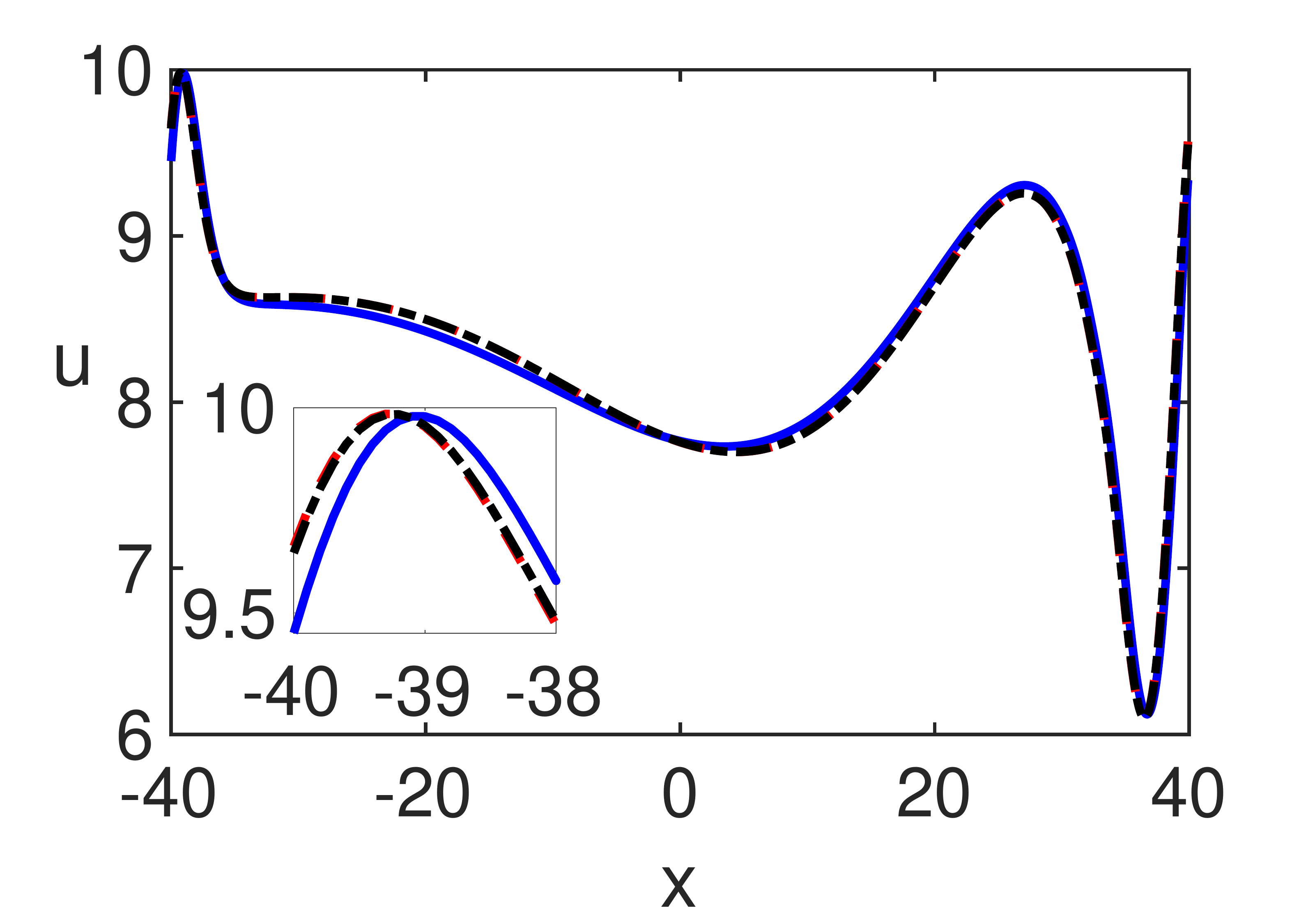}
	\caption{\small A comparison of the numerical solution of the BKG equation (blue, solid line) and the constructed semi-analytical solution including leading-order (red, dashed line) and $O \lb \sqrt{\varepsilon} \rb$ (black, dash-dotted line) terms, at $t=1/\varepsilon$. Parameters are $L=40$, $N=800$, $k = 1/\sqrt{3}$, $c = \alpha = \beta = \gamma = 2$, $\varepsilon = 0.001$, $\Delta t = 0.01$, $\Delta T = \varepsilon \Delta t$, $d_{1} = 5$ and $d_{2} = 0.5$. There is a good agreement between the numerical solution and the constructed solution.}
	\label{fig:FVSoliton}
\end{figure}

To illustrate the validity of the constructed solution we consider two examples. In both cases we will assume $c = \alpha = \beta = \gamma = 1$. Firstly let us consider a soliton on a raised pedestal, defined by the initial condition
\begin{align}
u(x,0) &= A \sechn{2}{\frac{x}{\Lambda}} + d_{1}, \\
u_t (x,0) &= \frac{2cA}{\Lambda} \sechn{2}{\frac{x}{\Lambda}} \tanhn{ }{\frac{x}{\Lambda}} + d_{2},
\label{ICFV}
\end{align}
where $d_{1}$, $d_{2}$ are constants and we have
\begin{equation}
A = \frac{6ck^2}{\alpha}, \quad \Lambda = \frac{\sqrt{2c \beta}}{k},
\label{BOstCoefsFV}
\end{equation}
with $k = \sqrt{\alpha/3c}$. We take $d_{1} = 5$, $d_{2} = 0.5$, $\varepsilon = 0.001$ and present the results at $t = 1/\varepsilon$ in Figure \ref{fig:FVSoliton}. We can see that there is a good agreement between the weakly-nonlinear solution and the results of direct numerical simulations, with a small phase shift between the constructed solution and the direct numerical solution. 

\begin{figure}[!htbp]
	\centering
	\includegraphics[width=0.8\textwidth]{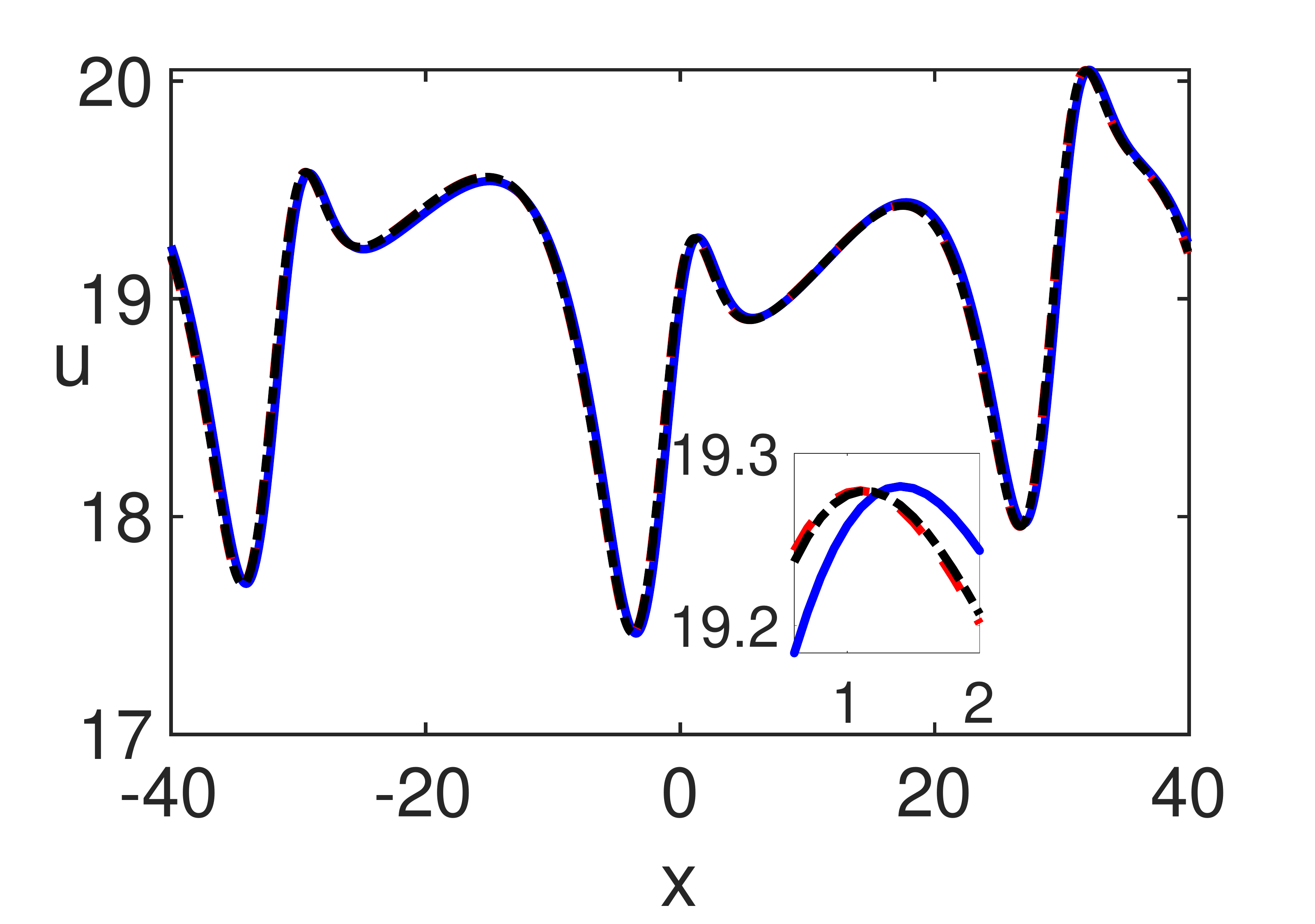}
	\caption{\small A comparison of the numerical solution of the BKG equation (blue, solid line) and the constructed semi-analytical solution including leading-order (red, dashed line) and $O \lb \sqrt{\varepsilon} \rb$ (black, dash-dotted line) terms, at $t=1/\varepsilon$, for a cnoidal wave initial condition. Parameters are $L=40$, $N=800$, $c = \alpha = \beta = \gamma = 1$, $\varepsilon = 5 \times 10^{-4}$, $\Delta t = 0.01$, $\Delta T = \varepsilon \Delta t$, $d_{1} = 5$ and $d_{2} = 0.5$. There is a good agreement between the numerical solution and the constructed solution.}
	\label{fig:FVCnoidal}
\end{figure}

The second example we consider is for a cnoidal wave initial condition. The exact cnoidal wave solution of the equation
\begin{equation*}
	2 c f_{T}^{-} + \alpha f^{-} f_{\xi_{-}}^{-} + \beta c^2 f_{\xi_{-} \xi_{-} \xi_{-}}^{-} = 0,
\end{equation*}
can be written in terms of the Jacobi elliptic function as follows (e.g., \cite{Johnson97})
\begin{align}
	&f^{-} =  - \frac{6 \beta c^3}{\alpha} \left (f_2 - (f_2 - f_3) \mathrm{cn}^2[(\xi^- + v T) \sqrt{\frac{f_1 - f_3}{2}} | m] \right ), \\
&\mbox{where} \quad v =  (f_1 + f_2 + f_3) \beta c^2, \quad m = \frac{f_2 - f_3}{f_1 - f_3}.
\end{align}
This solution is parametrised by the constants $f_3 < f_2 < f_1$ such that the elliptic modulus $0 < m < 1$. The wave length can be calculated as
\begin{equation*}
L = 2 K(m)  \sqrt{\frac{2}{f_1 - f_3}},
\end{equation*}
where $K(m)$ is the complete elliptic integral of the first kind. In our example we take $f_{1} = 2\times 10^{-5}$, $f_{2} = 0$, $f_{3} = -1/3$, giving $m \approx 0.999$. We take $c = \alpha = \beta = \gamma = 1$, $\varepsilon = 5 \times 10^{-4}$ and the same pedestal as for the soliton case i.e. $d_{1} = 5$ and $d_{2} = 0.5$. The results are presented at $t = 1/\varepsilon$ in Figure \ref{fig:FVCnoidal}. We can see that again there is a good agreement between the weakly-nonlinear solution and the results of direct numerical simulations, although the phase shift between the constructed solution and the direct numerical solution is larger in this case. The inclusion of higher-order terms (at $O\lb\varepsilon\rb$) improved the solution a little but more terms from the expansion are required for better agreement.

\section{Conclusions}
In this paper we discussed the scattering of long pure and radiating bulk strain solitary waves in a delaminated bi-layer and a related initial-value problem. The modelling was performed within the framework of the nondimensional Boussinesq-type equations. We highlighted key features of the behaviour of pure and radiating solitary waves in such delaminated bi-layers, which could be used for introscopy of layered structures, in addition to traditional tools. The fission of a single incident soliton into a group of solitons in the delaminated area of a perfectly bonded PMMA bi-layer has been observed in \cite{Dreiden10}. The generation of a radiating solitary wave and subsequent disappearance of the ``ripples" in the delaminated area of a two-layered PMMA bar with the PCP (polychloroprene-rubber-based) adhesive has been observed in  \cite{Dreiden12}.  Our numerical modelling motivates further laboratory experimentation with other materials used in applications. We also discussed how one can construct a weakly-nonlinear solution of the related initial-value problem  avoiding restrictions on the mass of the initial conditions, which opens the way to extending the semi-analytical approaches to the scattering of solitary waves discussed in this paper to the scattering of periodic waves.









%
%
%

\end{document}